# Learning of High Dengue Incidence with Clustering and FP-Growth Algorithm using WHO Historical Data

Franz Stewart V. Dizon
Institute of Information and Computing Sciences
University of Santo Tomas
Manila, Philippines
frdizon@live.com

Stephen Kyle R. Farinas
Institute of Information and Computing Sciences
University of Santo Tomas
Manila, Philippines
kylefarinas@gmail.com

Reynaldo John Tristan H. Mahinay Jr.
Institute of Information and Computing Sciences
University of Santo Tomas
Manila, Philippines
rjt.mahinay@gmail.com

Harry S. Pardo
Institute of Information and Computing Sciences
University of Santo Tomas
Manila, Philippines
harrypardo@outlook.ph

Cecil Jose A. Delfinado
Institute of Information and Computing Sciences
University of Santo Tomas
Manila, Philippines
cadelfinado@ust.edu.ph

*Abstract*— This paper applies FP-Growth algorithm in mining fuzzy association rules for a prediction system of dengue. The system mines its rules through input of historic predictor variables for dengue. The rules will be used to build a rule-based classifier to predict the dengue incidence for the next month for the years 2001-2006 in the Philippines. The FP-Growth Algorithm was compared to Apriori Algorithm by Sensitivity, Specificity, PPV, NPV, execution time and memory usage. The results showed that FP-Growth Algorithm is significantly better in execution time, numerically better in memory and comparable in Sensitivity, Specificity, PPV and NPV to Apriori Algorithm.

## I. INTRODUCTION

Dengue is one of the most prevalent disease in the Philippines. Early warning system could help the local government for outbreaks [7]. Today, machine learning is important in health and medicine since it process large datasets and give automated results [3].

Fuzzy Association Rule Mining to predict weekly dengue incidence was a method used by Buczak et al. This system Apriori Algorithm as the rule miner [2]. We used FP-Growth Algorithm instead as the rule miner since Han, Pei, Yiwen, Heaton, Mythili and Shanavas stated that it can solve the problem of Apriori Algorithm in generating frequent itemsets [4][5][6].

Hence, we compared the performance of Apriori and FP-Growth Algorithm in mining the rules for Fuzzy Association Rule Mining.

## II. RELATED STUDIES

### A. Fuzzy Association Rule Mining

In the study of "Prediction of High Dengue Incidence in the Philippines" by Buczak et al., it is stated that it is important that future high detection of disease (due to an outbreak) must be known. To create this prediction model, they used predictor variables such as historical epidemiological, environmental, socio-economic data and weather pattern and find association in it using Fuzzy Association Rule Mining to predict weekly dengue incidence in some specific provinces in the Philippines four weeks in advance. The results were accurate based on a performance metrics and can be modified in other diseases.

### B. Apriori Algorithm

Agrawal and Srikant created the Apriori Algorithm as a rule mining technique in a transactional database. It used join and prune process for finding frequent patterns and deriving association rules from them. The join process (candidate generation) is by joining k-itemsets and prune process for disregarding itemsets [1][5][6].

According to Jeff Heaton, Apriori Algorithm is computational expensive and has scalability issues [5]. Mythili and Shanavas stated in their study that the algorithm has large memory consumption and more time needed to run [6].

### C. FP-Growth Algorithm

FP-Growth Algorithm was introduced by Han, Pei and Yin in 2000 to eliminate the candidate generation of Apriori Algorithm. It uses "FP-Tree" to store the transaction of a database. It traverses the tree to form conditional fp-trees in a bottom-up approach [4][5][6]. Mythili and Shanavas said that it has less memory consumption because it utilizes the database into 2 scans only [6].

## III. METHODOLOGY

The main objective of the study is to compare the performance of Buczak et al.'s rule miner which is Apriori Algorithm to the proposed FP-Growth Algorithm. To prove that it is indeed true, we reconstructed the system and performed an experiment to test if the proposed rule miner is better. The materials used for constructing the algorithm, interpolation and documentation were Python and its libraries (Sklearn, SciPy, NumPy, Pandas, PyODBC), C#, Google Docs and Microsoft Live.

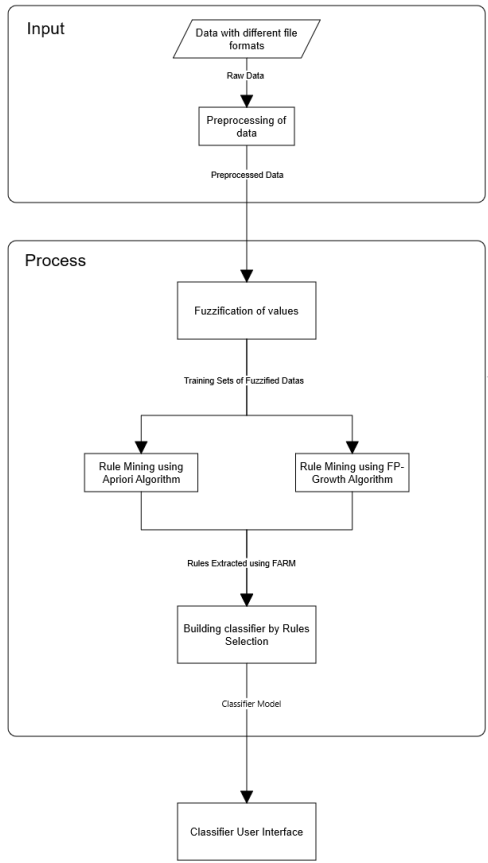

**Fig 1. System Architecture.**

There are four main components that comprises the whole system in Fig 1 namely, Preprocessing of Data, Fuzzification Module, Rule Mining and Classification.

*A. Sources of Data*

**Rainfall** - NASA Tropical Rainfall Measuring Mission
**Temperature** - USGS Land Processes Distributed Active Archive Center
**Typhoon Status and Wind** - Unisys Weather
**NDVI ( Normalized Difference Vegetation Index)**-
USGS Land Processes Distributed Active Archive Center
**EVI ( Enhanced Vegetation Index) -** USGS Land Processes Distributed Active Archive Center
**Southern Oscillation Index** –
US National Center for Atmospheric Research
**Sea Surface Temp Anomaly** - NASA Global Change Mastery Directory
**Population and Poverty** - Philippines National Statistic Office
**Political Stability** - Worldwide Governance Indicators Project
**Dengue Cases** - World Health Organization

The input of the system is features in different formats and dengue data as shown in Section III-A. To deal with different formats preprocessing and interpolation is needed to transform and fill-in the missing values of the data.

*B. Preprocessing of Data*

Different programs and libraries are needed to view and read other types of data. NDVI, EVI, rainfall and temperature have a format of hierarchical data format 4 (.hdf). For storing the data in the database, the .hdf file is converted to hierarchical data format 5 (.h5) and PyODBC. The other data were in .csv format and were easily read and manipulated. Transforming the .hdf file, the longitude and latitude of .hdf files were associated to the longitude and latitude of each region. Another transformation is measuring the typhoon distance to each region.

Gathered data have different temporal resolution with missing values which is needed to interpolate to monthly data. Using Pandas, it interpolates the data to monthly data and filling in the missing values.

*C. Fuzzification of Data*

The fuzzification module accepts input of data after the preprocess. It consists of functions such as normalizing the data, clustering by k-Means and fuzzy set membership function. The new values were determined using normalization.

Each normalized data per feature was used in k-Means clustering and fuzzy membership function. The number of cluster made correspond to the membership fuzzy sets. After clustering, each data point was fuzzified using trapezoidal and triangular functions.

```
KMeans(n_clusters, random_state=0){

Pick K points randomly (random_state = 0) as
initial cluster centroids (means) corresponding to
the number of n_clusters.

Repeat:
    Assign points to the nearest mean.
    Assign each mean to the average of its
    assigned points.
    Stop repeat if no point changes cluster
}
```
**Algorithm 1. k-Means Algorithm**

The input of Algorithm 1. is the number of clusters (n_clusters), which is 4 for the features and 2 for dengue data. The algorithm picks random points as the first 4 cluster for features and first 2 for dengue data. It repeatedly assigns new clusters by averaging the points assigned to it until the points doesn't change to a new cluster. After clustering, the points now have their respective clusters.

The output is a fuzzified data which is the conversion of the points to membership fuzzy sets based on the triangular and trapezoidal function using the clusters.

*D. Rule Mining*

Buczak et al. uses Apriori Algorithm and we used FP-Growth Algorithm instead for mining the rules. The rules of the 2 algorithms will be used to build a rule-based classifier and compare their performance.

```
FPGrowth(FPTree, a, support) {
    for each item a_i in the header of FPTree {
        generate β = a_i ∪ FPTree with support = a_i.support
        construct β conditional pattern base and conditional FP-Tree (Tree β)
        if Tree β != null
            FP-Growth (FPTree β, β)
    }
    return frequent_patterns(FPTree)
}
```
**Algorithm 2. FP-Growth Algorithm**

The input of Algorithm 2 is the fuzzified data from the fuzzification module. The FPTree is preconstructed by scanning the database in finding frequent items (fuzzified data) in descending support count. Insertion of items depends if tree β has child N such that N.item = previous item, then count is increased. Else, tree β create new node and link to parent β. Conditional pattern base is a path of the tree which an item occurs. Construction of conditional FPTree is finding similar item in pattern base and sum up the its count. Mining the frequent patterns is combining the item and conditional FPTree.

The output is a list of association rules using the frequent patterns and given minimum confidence.

*E. Classification*

In constructing the rule-based classifier, the rules from Apriori and FP-Growth Algorithm were used. The rules were sorted in consequent, confidence, number of antecedents and lift (for Apriori). The test data will be iterated to classify the next month dengue class. This will be done both for Apriori and FP-Growth Algorithm. The features of the data will be the antecedent and a matching rule with the same antecedent will output a consequent (Dengue Next_High and Dengue Next_Low). The consequent will be compared to the test data's data class to verify accuracy.

IV. ANALYSIS AND INTERPRETATION OF RESULTS

*A. Test Results*

The dengue data used in the system are in the years of 2001 to 2006. The features and dengue data were sampled using systematic sampling. We partitioned the data by 80% training set and 20% test set.

Table 3 shows the average execution time and memory usage of the algorithms for the test data. FP-Growth is higher than Apriori in these metrics.

**Table 3. Execution Time and Memory Usage Comparison.**

|  | Apriori | FP-Growth |
|---|---|---|
| Memory Usage | 992.8 MiB | 106.4 MiB |
| Execution Time | 2753.2999 sec | 20.3923 sec |

Table 4 presents the overall True Positives (TP) and True Negatives (TN) where TP is the correct prediction for high incidence and TN for the correct prediction of low incidence. Overall False Positive (FP) and False Negatives (FN) for incorrect high incidence prediction and incorrect low incidence prediction for the 13 regions in the Philippines.

**Table 4. True Positives, False Positives, True Negatives, False Negatives.**

| Algorithm | True Positives | False Positives | True Negatives | False Negatives |
|---|---|---|---|---|
| Apriori | 14 | 1 | 139 | 12 |
| FP-Growth | 13 | 1 | 139 | 13 |

Table 5 shows the computed accuracy metrics based from the results shown in Table 3. As stated by Buczak et al., these metrics were used to evaluate their system.

- **Sensitivity**= $\frac{TP}{TP+FN}$ which is the proportion of correctly predicted outbreaks.
- **Specificity**= $\frac{TN}{TN+FP}$ which is the proportion of correctly predicted non-outbreaks.
- **Positive Predictive Value (PPV)**= $\frac{TP}{TP+FP}$ which is the positive prediction that are outbreaks.
- **Negative Predictive Value (NPV)**= $\frac{TN}{TN+FN}$ which is the negative prediction that are non-outbreaks.

Sensitivity and PPV are inversely related to some extent, if the number of false positive increases for PPV then the number of false negative decreases in sensitivity and vice versa. These two metrics were formed to a single value with a score of 1 to balance both metrics.

- $F_1$ **Measure**= $2(\frac{Sensitivity*PPV}{Sensitivity + PPV})$

**Table 5. Accuracy Metrics Table.**

|  | Apriori (%) | FP-Growth (%) |
|---|---|---|
| Sensitivity | 53.85 | 50.00 |
| Specificity | 99.29 | 99.29 |
| PPV | 93.33 | 92.86 |
| NPV | 92.05 | 91.45 |
| $F_1$ Measure | 68.29 | 65.00 |

*B. Analysis of Results*

The hypothesis for this test was stated as follows: The first claim was the execution time of FP-Growth Algorithm was significant than Apriori Algorithm. The P-value was less than the critical value the decision is to reject the null hypothesis and the difference is significant.

The result of both algorithms in memory usage can be observed that FP-Growth Algorithm consumed less memory than Apriori Algorithm in mining the frequent itemsets.

The second claim was there is no significant difference between FP-Growth Algorithm and Apriori Algorithm in predicting dengue incidence.

In Specificity both algorithms were good in correctly predicting low incidence. However, Apriori Algorithm is higher than FP-Growth Algorithm in Sensitivity. The test data were composed mostly of low dengue incidence. Thus, the rules mined is not good enough for high dengue incidence. The test value was less than the critical value and so we cannot reject the null hypothesis

In PPV and NPV, Apriori Algorithm usually predict that high incidence may occur than FP-Growth Algorithm. The test value also was less than the critical and so we cannot reject the null hypothesis.

The F-measure used a score of 1 to balance Sensitivity and PPV. We can say the FP-Growth Algorithm is numerically and insignificantly lower than Apriori Algorithm.

## V. SUMMARY, CONCLUSIONS AND RECOMMENDATIONS

*A. Summary*

Our study was about finding a way to improve a current model in dengue prediction. The researchers used the rule-mining technique FP-Growth Algorithm instead of Apriori Algorithm. The accuracy metrics, execution time and memory usage of the two algorithms were compared against each other. The system was tested using the dengue data from the World Health Organization from the years, 2001 to 2006 for the 13 regions in the Philippines.

The model was evaluated by its accuracy metrics, execution time and memory usage. The rule-mining technique, Apriori Algorithm was created to use for comparison. Statistical treatment of data was used specifically, z-test for the difference between two proportions for the accuracy metrics and P-Value method for execution time, to test the hypothesis of the study. However, evaluating in terms of memory usage, FP-Growth Algorithm was numerically compared to Apriori Algorithm.

After evaluating the model using appropriate statistical treatments and a numerical comparison for memory usage, we can now say which of the two algorithms is more efficient in rule mining for dengue prediction.

*B. Conclusions*

The main objective of the study was to improve the rule mining of Fuzzy Association Rule Mining for prediction of high dengue incidence in the Philippines. Using statistical treatments and tests to evaluate the performance of the rule miners, we can now conclude our findings to the corresponding problem.

Based on the results. The execution time and memory usage of FP-Growth Algorithm were 20.3923 secs and 106.4 MiB while Apriori Algorithm obtained 2753.2999 secs and 992.8 MiB. Using the P-Value method for execution time and comparing the numerical results in memory usage, it can be concluded that FP-Growth Algorithm is significantly better than Apriori Algorithm in time and numerically better in memory consumption.

The accuracy metric results of FP-Growth Algorithm were Sensitivity = 0.5, Specificity = 0.9929, PPV = 0.9286, NPV = 0.9145 and $F_1$ = 0.65 while Apriori Algorithm obtained Sensitivity = 0.5385, Specificity = 0.9929, PPV = 0.9333, NPV = 0.9205 and $F_1$ = 0.6829. Using hypothesis for z test for the difference between two proportions, it can be concluded that there is no significant difference between FP-Growth Algorithm and Apriori Algorithm.

The main objective of our study was accomplished because the execution time had significant results and comparable accuracy metrics given the limitations of the data.

*C. Recommendations*

Based on the conclusions, we experienced some difficulties and the system has several areas of improvement. The addition of dengue cases data is recommended since it affects how the algorithm learns. Use other hybrid of algorithms rather than the Fuzzy Association Rule Mining for more accurate classification of high dengue incidence. Use data from other countries to test the performance of the algorithms.

ACKNOWLEDGMENT

The researchers would like to thank the following who made this study possible. Ms. Ria Sagum for her suggestions, Mr. Khrisnamonte Balmeo, Ms. Cherry Rose Estabillo and Mr. Jonathan Cabero for providing us information to improve in our system.